\documentclass[preprint]{aastex}

\RequirePackage{epstopdf}

\bibpunct{(}{)}{,}{a}{}{,}

\begin{document}
\received{2012 January 6}
\revised{2012 January 30}
\accepted{  }
\slugcomment{Accepted by ApJ Letters}

\shortauthors{Adams et al. }
\shorttitle{Properties of Protostars and Disks in OMC-2 with SOFIA/FORCAST}

\title{First Science Observations with SOFIA/FORCAST: Properties of Intermediate-Luminosity Protostars and Circumstellar Disks in OMC-2}

\author{Joseph D. Adams\altaffilmark{1}}
\author{Terry L. Herter\altaffilmark{1}}
\author{Mayra Osorio\altaffilmark{2}}
\author{Enrique Macias\altaffilmark{2}}
\author{S. Thomas Megeath\altaffilmark{3}}
\author{William J. Fischer\altaffilmark{3}}
\author{Babar Ali\altaffilmark{4}}
\author{Nuria Calvet\altaffilmark{6}}
\author{Paola D'Alessio\altaffilmark{7}}
\author{James M. De Buizer\altaffilmark{8}}
\author{George E. Gull\altaffilmark{1}}
\author{Charles P. Henderson\altaffilmark{1}}
\author{Luke D. Keller\altaffilmark{9}}
\author{Mark R. Morris\altaffilmark{5}}
\author{Ian S. Remming\altaffilmark{10}}
\author{Justin Schoenwald\altaffilmark{1}}
\author{Ralph Y. Shuping\altaffilmark{8}}
\author{Gordon Stacey\altaffilmark{1}}
\author{Thomas Stanke\altaffilmark{11}}
\author{Amelia Stutz\altaffilmark{12}}
\author{William Vacca\altaffilmark{8}}

\altaffiltext{1}{Cornell University, Department of Astronomy, Space Sciences Bldg., Ithaca, NY, USA 14853}
\altaffiltext{2}{Instituto de Astrofisica de Andalucia, CSIC, Camino Bajo de Hu\'{e}tor 50, E-18008 Granada, Spain}
\altaffiltext{3}{University of Toledo, Department of Physics and Astronomy, Mailstop 111, 2801 West Bancroft Street, Toledo, Ohio 43606, USA }
\altaffiltext{4}{NHSC/IPAC/Caltech, 770 South Wilson Avenue, Pasadena, CA 91125, USA}
\altaffiltext{5}{University of California, Los Angeles, Department of Physics and Astronomy, 405 Hilgard Ave., Los Angeles, CA 90095-1547, USA}
\altaffiltext{6}{Department of Astronomy, University of Michigan, 825 Dennison Building, 500 Church St, Ann Arbor, MI 48109, USA}
\altaffiltext{7}{Centro de Radioastronomia y Astrofisica, Universidad Nacional Autonoma de Mexico, 58089 Morelia, Michoacan, Mexico}
\altaffiltext{8}{SOFIA-University Space Research Association, NASA Ames Reseach Center, Mail Stop N211-3, Moffett Field, CA 94035, USA}
\altaffiltext{9}{Ithaca College, Physics Department, 264 Ctr for Natural Sciences, Ithaca, NY 14850, USA}
\altaffiltext{10}{Department of Physics and Astronomy, University of Rochester, Rochester, NY 14627, USA}
\altaffiltext{11}{ESO, Karl-Schwarzschild-Strasse 2, 85748 Garching bei M\"{u}nchen, Germany}
\altaffiltext{12}{Max-Planck-Institut f\"{u}r Astronomie, K\"{o}nigstuhl 17, 69117 Heidelberg, Germany}

\begin{abstract}
We examine eight young stellar objects in the OMC-2 star forming region based on observations from the SOFIA/FORCAST early science phase, the {\it Spitzer Space Telescope}, the {\it Herschel Space Observatory}, 2MASS, APEX, and other results in the literature. We show the spectral energy distributions of these objects from near-infrared to millimeter wavelengths, and compare the SEDs with those of sheet collapse models of protostars and circumstellar disks. Four of the objects can be modelled as protostars with infalling envelopes, two as young stars surrounded by disks, and the remaining two objects have double-peaked SEDs. We model the double-peaked sources as binaries containing a young star with a disk and a protostar.  The six most luminous sources are found in a dense group within a $0.15 \times 0.25$ pc region; these sources have luminosities ranging from 300 $L_\odot$ to 
20 $L_\odot$.  The most embedded source (OMC-2 FIR 4) can be fit by a class 0 protostar model having a luminosity of $\sim 50~L_\odot$  and mass infall rate of
$\sim 10^{-4}~M_\odot~\rm{yr}^{-1}$. 
\end{abstract}


\section{INTRODUCTION}
OMC-2 is a star forming region embedded in the molecular filament stretching northward of the Orion Nebula and is one of the most 
luminous regions in the Orion A molecular cloud. Recent distance estimates place it at $\sim 420$ pc
\citep{hirota07,jeffries07,menten07,sandstrom07}. As a site of star formation activity, it contains a large number of protostars and
pre-main sequence stars with dusty circumstellar disks \citep{peterson08}. Therefore, OMC-2 represents a laboratory for the investigation of 
the detailed properties of protostars and disks which are forming in an active environment.

\citet{gatley74} first identified OMC-2 as a complex of pre-main sequence objects associated with nearby CO emission. 
\citet{thronson78} measured the infrared luminosity of OMC-2 to be 2100 $L_\odot$, but did not resolve 
the far-IR components of IRS 3 and IRS 4 and thus could not provide a measurement of their individual SEDs.
Later work expanded the coverage of OMC-2 to higher resolution at the longer wavelengths. \citet{chini97} conducted a millimeter study
that resolved several cold (20 K) dusty cores throughout OMC-2/3. In OMC-2, these cores are associated with submillimeter continuum emission \citep{lis98}. 
The detection of centimeter sources with the VLA established many of the cores as bona fide protostars \citep{reipurth99}.
Recent studies have included high resolution, ground-based observations to further examine both the stellar component and the dusty cores \citep{nielbock03,shimajiri08}.
These studies revealed several stellar sources associated with the dusty cores and substructure in the cold dust of FIR 4.

While previous studies include high resolution observations at 10 and 20 $\mu$m, submillimeter, and millimeter wavelengths, there is a lack of data at high 
spatial resolution in the $\sim 20 - 200$ $\mu$m range which is critical
for constraining models of protostellar envelopes and for characterizing any external heating of both dusty 
disks and envelopes. During its early science phase, SOFIA \citep{young12} observed the Orion Nebula Cluster and OMC-2
with the FORCAST mid-IR camera \citep{adams10, herter12} and detected eight intermediate-luminosity protostars and disks in OMC-2. We combine these data with data from the 
2MASS near-infrared survey \citep{skrutskie06}, the {\it Spitzer}
Orion Molecular Cloud Survey \citep{megeath05}, the {\it Herschel} Orion Protostar Survey (HOPS; Fischer et al. \citeyear{fis10}), submillimeter data from APEX, 
and other published data in the literature to produce complete 
SEDs of these objects from near-IR to millimeter wavelengths. Finally, we compare the SEDs to those of model protostars and disks in order to 
derive the physical properties of these objects. 

\section{OBSERVATIONS}\label{sec:observations}
\subsection{SOFIA/FORCAST Observations}
SOFIA/FORCAST observed OMC-2 during the SOFIA short science phase on December 1 and December 8, 2010, at an altitude of $\sim 13100$ m.
Two pointings were used: ($05^h$ $35^m$, $-05^\circ$ $08^\prime$) and ($05^h$ $35^m$,  $-05^\circ$ $15^\prime$). Each pointing covered $3.4^\prime \times 3.2^\prime$,
with a (rectified) pixel scale of $0.77^{\prime\prime}$/pixel. The first pointing was observed at 19.7 $\mu$m and 37.1 $\mu$m (with the dichroic beamsplitter). The
second pointing was observed at 19.7 $\mu$m and 31.4 $\mu$m (with the dichroic), and 37.1 $\mu$m (without the dichroic). Observing modes C2N and C2NC2 \citep{herter12}
were used. A large ($4.8^\prime$) asymmetric chop throw was used to chop off nearby nebulosity and minimize off-axis coma in the on-source beam.
The chop frequency was $2 - 5$ Hz and the nod amplitude was approximately $20^\prime$. 
A 5-point, $10^{\prime\prime}$ step dither pattern was performed to reject
bad pixels during post-processing. The integration time in each dither position was approximately 30 sec. 
 
The data were processed with the reduction steps described in \citet{herter12}.
Point source photometry was performed on sources in the images using two-dimensional Gaussian fits to their radial brightness profiles. Typical
FWHM image quality values were $3.3^{\prime\prime}$ and $4.3^{\prime\prime}$ at 19.7 and 37.1 $\mu$m, respectively. During the December 8 flight, 
some elongation of the PSF was seen in the cross-elevation direction, resulting in image quality of 
$5.9^{\prime\prime}$ and $4.2^{\prime\prime}$ in the cross-elevation and elevation direction, respectively, in the second pointing field at 31.4 $\mu$m.
Photometric signal-to-noise was consistent with the mean point source sensitivity over 3 short science flights \citep{herter12}.

The extracted photometric fluxes were calibrated to a flux density using a standard instrument response derived from measurements of standard stars 
and solar system objects over several short science and basic science
flights \citep{herter12}. We applied small ($\lesssim 2\%$) color corrections to each source, since the
standard instrument response represents the response to a flat spectrum ($\nu F \nu = $ constant) source.
The estimated $3~\sigma$ uncertainty in the
calibration due to variations in flat field, water vapor burden, and altitude
is approximately $\pm 20\%$.

\subsection{{\it Spitzer} Observations}
{\it Spitzer}/IRAC images were taken in two epochs in 2004 March 09 and October 10 as part of program PID 43.  The reported photometry is the median value obtained from 
four 10.4 second high dynamic range mode frames.  
The magnitudes of the sources were extracted with aperture photometry from the Basic Calibrated Data using the IDL aper.pro 
program \citep{landsman93}.
A 2 pixel radius and 2 to 6 pixel background annulus was used (Megeath et al. in prep.), and a standard calibration was adopted (see Gutermuth et al. \citeyear{gutermuth09} and Reach et al. \citeyear{reach05}).  The MIPS scan map was made in 2004 March 20  with the fast rate and $160''$ cross scan offset as part of program PID 58, resulting in 30 seconds of integration time per pixel. The data were reduced with the MIPS instrument team's Data Analysis Tool (Gordon et al. 2005), and photometry was obtained using a the IDL implementation of DAOPHOT \citep{landsman93}. DAOPHOT was modified to ignore saturated pixels (Kruykova et al. submitted).  The calibration of \citet{engelbracht07} was adopted.  The magnitudes were computed to flux densities using the zero fluxes given in \citet{reach05} for IRAC and \citet{engelbracht07}
for MIPS.  Given the relatively insensitivity of the fluxes to the spectral shape, color corrections were not applied. 

{\it Spitzer}/IRS spectra were acquired for two sources (MIR 29 and MIR 31+32; Nielobock et al. \citeyear{nielbock03}) in the SOFIA/FORCAST fields-of-view. 
These $5 - 36$ $\mu$m spectra were obtained with the short-low and long-low modules (PIDs 30859 and 50734). 
We follow the method described in \citet{poteet11} to extract and calibrate the spectra.

\subsection{{\it Herschel} Observations}
The {\it Herschel Space Observatory} observed an $8^\prime$ square field encompassing HOPS 59, 60, 108,
368, 369, and 370 on 2010 September 28 (observing day 502, observation
IDs 1342205228 and 1342205229) in the 70 $\mu$m and 160 $\mu$m bands of the PACS instrument \citep{pog10} as part of the HOPS
Open Time Key Program. These data have angular resolutions of $5.2^{\prime\prime}$ and $12^{\prime\prime}$, respectively.  The target field was observed with 
homogeneous coverage using two orthogonal scanning directions and a scan speed of $20''$/s, with each scan repeated 8 times for a total
observation time of 3290~s.  To facilitate point source photometry, a high-pass filter was applied to the data to minimize the effects of drift \citep{fis10}.

The aperture photometry was obtained using a $9.6^{\prime\prime}$ aperture at 70 $\mu$m and a $12.8^{\prime\prime}$ aperture at 160 $\mu$m  with
a background annulus extending from the aperture limit to twice that value in both channels.   The results were corrected with the encircled energy fraction provided by the PACS consortium (priv. comm.).   The error estimate, which is dominated by calibration uncertainties, is estimated to be 5\% at
both wavelengths.

\subsection{APEX Data}
We include 350 and 870 $\mu$m photometry acquired at APEX with SABOCA and LABOCA, respectively (proposal ID 088.C-0994). 
\citet{stanke10} describe the observing and data reduction procedures for the APEX data.
The beam sizes of the final reduced maps are $7.3''$ and $19''$ for the SABOCA and LABOCA observations respectively. We  measured the flux density per beam at the position of the
source initially using the coordinates from the HOPS survey 70 $\mu$m point source photometry. 
If the source is well--detected, we re-centered the coordinates and measure the flux centered on the source in order to account for possible pointing offsets between data sets.
Otherwise, we adopt the PACS 70 $\mu$m coordinates and measure a flux which we can consider
an upper limit on the source flux at the wavelength of interest.  The errors are dominated by the calibration uncertainty and are therefore adopted to be 40\% of the measured 
flux at both wavelengths.  

\section{RESULTS}\label{sec:results}
\begin{figure}
\plotone{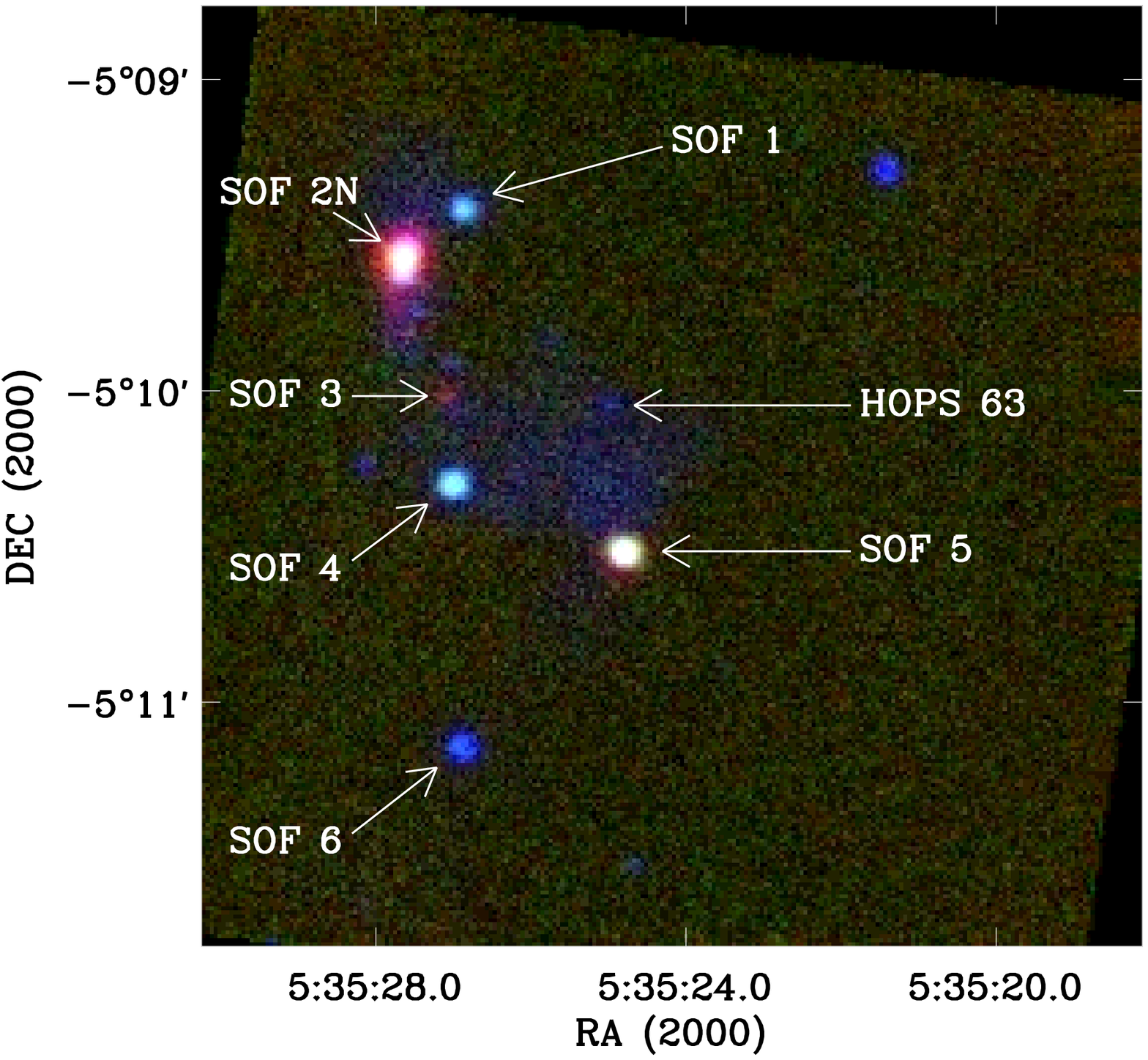}
\caption{False color image showing the northern SOFIA/FORCAST pointing field in OMC-2 at 4.5 $\mu$m (blue, IRAC), 19.7 $\mu$m (green, FORCAST), and 37.1 $\mu$m (red, FORCAST). SOFIA/FORCAST sources 1-6 are labeled; see Table \ref{tab:ids} for previous identifications. The protostellar candidate HOPS 63 was not detected by SOFIA/FORCAST. The remaining background point sources seen at 4.5 $\mu$m are dominated by photospheres and young stars with disks.
\label{fig:images}}
\end{figure}

\begin{figure}
\plotone{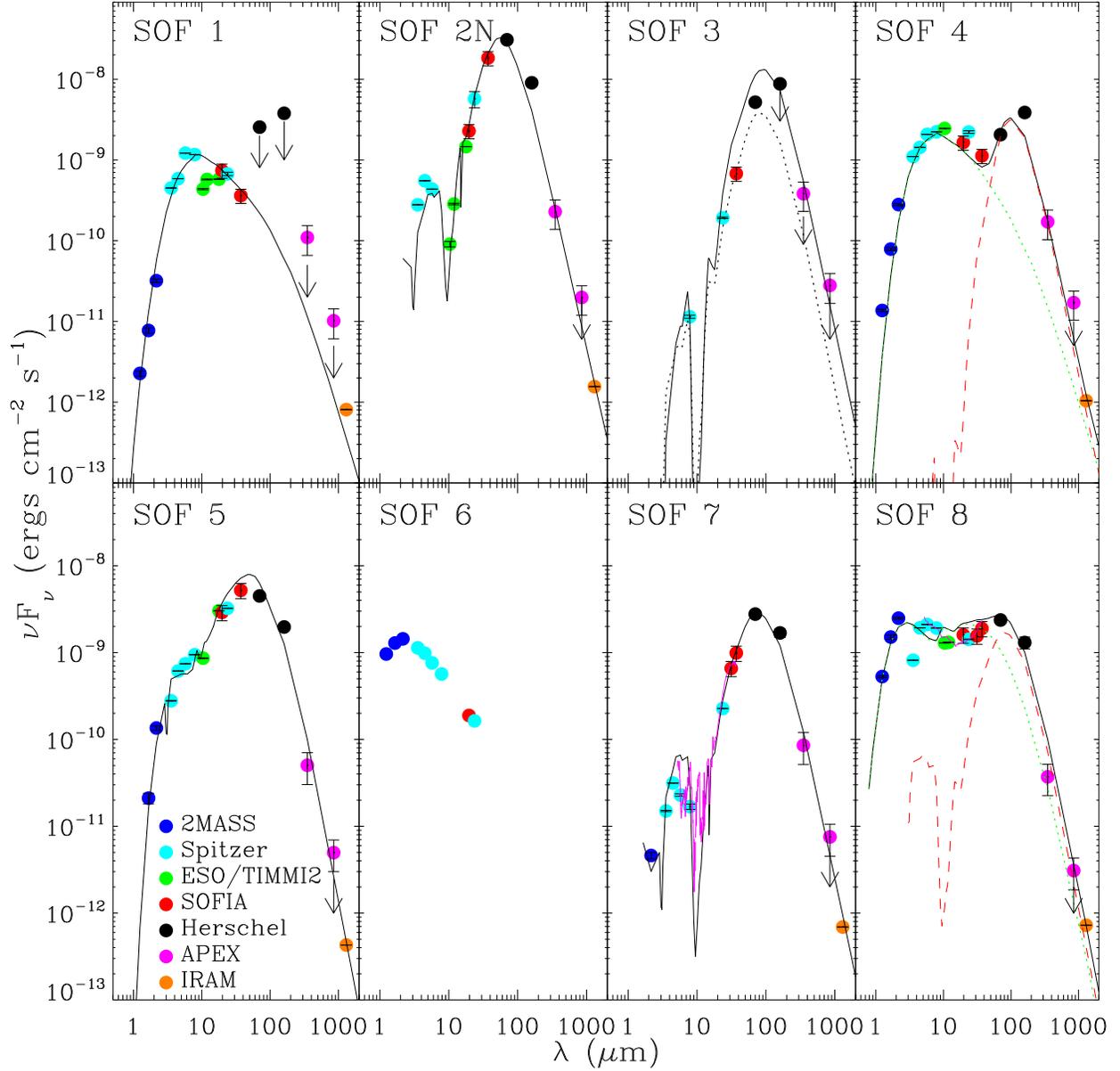}
\caption{SEDs of SOFIA/FORCAST sources SOF 1 - 8. The data points are color coded
by data set (facility). Uncertainties are shown with error bars and downward arrows indicate data points that are upper limits.  
{\it Solid black curves}: Total SEDs for the best-fit models. {\it Dotted green curves}: Disk component of binary systems. 
{\it Dashed red curves}: Envelope-dominated component of binary systems. {\it Dotted black curve}: Lower luminosity
model fit for SOF 3. Note that the absorption feature at 9.7 $\mu$m of source 
SOF 7 shows some emission, likely because of contamination by nearby sources. Only the red side of the feature has been fitted.
The SED shortward of $11~\mu$m is dominated by noise and has not been included in the fit. 
\label{fig:seds}}
\end{figure}

Fig. \ref{fig:images} shows a {\it Spitzer}/IRAC and SOFIA/FORCAST multi-wavelength false color image of the northern FORCAST field. In both FORCAST fields,
SOFIA/FORCAST detected most point sources (except the faint source HOPS 63, a likely protostar, and HOPS 64, a likely young stellar object), that were present in the
{\it Herschel}/PACS 70 and 160 $\mu$m images, plus the source SOF 6. This field shows a relatively compact group of luminous young stellar objects, 
with seven sources found in a $0.13 \times 0.25$ pc region.  The two remaining
sources, SOF 7 and SOF 8 are found to the south of the group and are not shown.
In Table \ref{tab:ids}, we list the eight SOFIA/FORCAST sources detected in the field and cross-correlate this list with previous identifications. 
In Table \ref{tab:fluxes}, we present the flux densities 
for each source. We used these flux densities to construct a SED for each source (Fig. \ref{fig:seds}), with additional values at
1-2 $\mu$m (2MASS), 10.4 $\mu$m, 11.9 $\mu$m, and 17.8 $\mu$m (Nielbock et al. \citeyear{nielbock03}), and 1.3 mm \citep{chini97}.
Fig. \ref{fig:seds} also shows that the {\it Spitzer}/IRS spectra for SOF 7 and SOF 8 are in excellent agreement with the FORCAST data.

\section{MODELS}

The physical properties of the SOFIA-detected objects were derived using
models consisting of an infalling envelope and/or a star+disk system.
For the envelope, we adopted the sheet collapse models developed by
Hartmann, Calvet \& Boss (\citeyear{hartmann96}, hereafter HCB96), which seem appropriate
given the filamentary nature of the OMC-2 molecular emission. In these
models, the envelope is flattened not only in the inner region, because
of rotation, but also at large scales due to the natural elongation of
the cloud. The inner region of the envelope is also partially excavated
by an outflow cavity. We adopted a dust opacity law that is based on the
observed class I object L1551 IRS5 \citep{osorio03}. The dust mixture includes
graphites, silicates, troilites and water ice compounds with a standard
grain-size distribution of $n(a) \propto a^{-3.5}$, with a minimum and
maximun size dust grains of 0.005 $\mu$m and 0.3 $\mu$m, respectively.
The temperature distributions and SEDs were calculated using radiative
codes developed by \citet{kenyon93} and by HCB96. 
In some cases, external heating of the envelope was added by using a single temperature 
modified blackbody \citep{poteet11} of temperature $\sim 30$ K; this is expected
in the OMC-2 region which is near the Orion Nebula and is likely 
heated by the massive stars in that regions.  

For the disk component of the model, we used the flared accretion disk models developed by
\citet{dalessio99,dalessio06}. In these models the disk density is given
by the conservation of angular momentum flux and an uniform mass
accretion rate. The temperature distribution is
determined from heating by viscous dissipation and stellar irradiation.
The disk radius is assumed to coincide with the centrifugal radius
(which is the largest radius on the equatorial plane that
receives the infalling material). The disk models also
account for dust evolution given by grain growth and settling towards
the disk midplane. We assumed that the dust composition and grain-size
distribution is the same as in the envelope, except in the disk midplane
where a maximum grain size of 1 mm is adopted.

The fits were found by visual inspection, following the SED behaviour as
a function of different physical parameters given in \citet{debuizer05} 
and \citet{dalessio99,dalessio06}. In some cases, the
inclination angle could be constrained from available near-IR images.

\section{DISCUSSION OF OBJECTS}

The best-fitting model SEDs are overlayed with the data in Fig. \ref{fig:seds}. The constrained model parameters for seven objects are 
listed in Table \ref{tab:modelpars}. We provide comments on each object:\\

\noindent SOF 1: The $\lambda >70$ $\mu$m fluxes of this source are considered 
upper limits due to contamination from extended emission associated with 
SOF 2N. Therefore, we fit this object by taking into account only the SED at $\lambda < 70$ $\mu$m.
The best fitting model consists of a young star surrounded by a disk with a
high mass accretion rate ($5 \times10^{-6}$ $M_\odot$ yr$^{-1}$). \\

\noindent SOF 2N: \citet{cohen77} presented near-IR images showing that IRS 4 has two 
distinct components, which are flagged as confused in the 2MASS catalog. 
Both components (MIR 21 and MIR 22) were resolved 
by Nielbock et al. (\citeyear{nielbock03}) at 10.4 and 17.8 $\mu$m. We
assume the northern component emits most of the flux at $\lambda \ge 17.8$ $\mu$m,
as evidenced by the relative flux densities at $17.8~\mu$m, and this is the component we model.
A high luminosity of 300 $L_\odot$ as well as some external heating is 
required by our model to explain the SOFIA flux at 37 $\mu$m.  
Therefore, SOF 2N likely hosts an intermediate-mass protostar.
A highly flattened, nearly edge-on envelope with a large opening-angle cavity is also required to 
reproduce the short wavelength data. \\

\noindent SOF 3: This deeply embedded source is a class 0 protostar associated with FIR 4 and was not detected by Nielbock et al. (2003) 
at 10.4 $\mu$m. This object has a large envelope mass (2.5 - 10 $M_\odot$) and  and dominates the submillimeter emission in this field. 
Two fits were obtained: one which treated the 160 micron as a detection, and one which treated 160 um as an upper limit 
(which would be the case if the 160 um emission were contaminated by the extended emission).  The resulting
luminosites range $30-50~L_\odot$ and the implied infall rates range from $4 \times 10^{-4}$ to $1 \times 10^{-4}~M_\odot~{\rm yr}^{-1}$; 
the highest infall rates in our sample (see Table 3). \citet{shimajiri08} resolved several small dust clumps toward FIR 4 at 
3.3 mm, but none of these clumps shows a peak that is clearly coincident with our $8-160~\mu$m detections.

\noindent SOF 4: The SOFIA data reveal that the SED has a double peak.
Our attempts to model this source as a single object
failed to explain the two peaks. Rather, we interpret this source as a
double stellar system, where one of the stars (more evolved)
is surrounded only by the disk component that dominates the emission at
$\lambda < 40$ $\mu$m. The other star (likely in an earlier
evolutionary phase) is deeply embedded into its infalling
envelope that dominates the emission at $\lambda > 40$ $\mu$m. The best
fit to the disk source requires a high mass accretion rate 
($\dot M_{acc}=10^{-5}~M_{\odot}~\rm{yr}^{-1}$). \\

\noindent SOF 5: Bipolar nebulosity from this object is seen at $3 - 8~\mu$m (Megeath et al., in prep.). 
This can be used to constrain the inclination angle, resulting in a better estimate 
of the luminosity and the density inferred from the modeling.\\

\noindent SOF 6: This is a young stellar object with a disk. It was not detected at $\lambda \ge 30~\mu$m and we do not
model this source. \\

\noindent SOF 7: The model for this object requires a central luminosity of
15 $L_\odot$ and a contribution from external heating. A {\it
Spitzer}/IRS spectrum shows a deep silicate absorption at 9.7 $\mu$m
which is fitted by assuming a high inclination angle and a large cavity.
Inspection of the 3.6 and 4.5 $\mu$m images shows a small lobe of
nebulosity extending $\sim 4''$ to the northeast. \\

\noindent SOF 8: Nielbock et al. (\citeyear{nielbock03}) resolved a double 
source (MIR 31 and 32) with projected separation $2.4''$ at this location at 10.4 and 11.9 $\mu$m. This 
object has a nearly flat SED from $\sim 2~\mu$m to 160 $\mu$m. We 
achieved a best fit to the SED using a binary system model. The best fit 
model indicates that in one of the sources the emission arises from a 
disk around an intermediate mass star and produces a luminosity of $130~L_\odot$.
The companion was fit by a protostar model that contains a dense envelope of $3~M_\odot$ 
irradiated by a central luminosity of $7~L_\odot$. \\

\section{CONCLUSION}
We have used SOFIA early science observations to examine the detailed properties of eight intermediate-luminosity ($7 - 300~L_\odot$)
protostars and circumstellar disks in OMC-2. Based on the model fits, it is believed that four of these objects are protostars and 
two are young stars with circumstellar disks, while two are likely young binary systems.
This work demonstrates the potential of SOFIA to observe protostars in active, luminous star forming regions, while simultaneously complementing 
both space- and ground-based observations. With such observations, we can characterize in detail the properties
of protostars and examine the dependence of star formation on environment.

\acknowledgments
We thank R. Grashius, S. Adams, H. Jakob, A. Reinacher, and U. Lampeter for their SOFIA telescope engineering and operations support. We also thank
the SOFIA flight crews and mission operations team (A. Meyer, N. McKown, C. Kaminski) for their SOFIA flight planning and flight support.
This work is based on observations made with the NASA/DLR Stratospheric Observatory for Infrared Astronomy (SOFIA). SOFIA science mission operations are conducted jointly by the Universities Space Research Association, Inc. (USRA), under NASA contract NAS2-97001, and the Deutsches SOFIA Institut (DSI) under DLR contract 50 OK 0901. Financial support for FORCAST was provided to Cornell by NASA through award 8500-98-014 issued by USRA.
M.O. acknowledges support from MICINN (Spain) grants AYA2008-06189-C03 
and AYA2011-30228-C03-01 (co-funded with FEDER funds).
This work is based on observations made with the {\it Spitzer Space Telescope}, which is operated by JPL/Caltech under NASA contract 1407. 

{\facility {\it Facilities}: Spitzer, SOFIA, Herschel, APEX}

\newpage

\newpage

\newpage
\begin{deluxetable}{lccccccccc}
\tabletypesize{\small}
\tablecaption{\label{tab:ids}Positions and identifications for all sources detected by SOFIA/FORCAST (SOF). IRS IDs are given in \citet{gatley74}, MIR IDs are given in Nielbock et al. (\citeyear{nielbock03}), FIR IDs are given in \citet{chini97} and \citet{shimajiri08}, CSO IDs are given in \citet{lis98}, and VLA IDs are given in Reipurth et al. (\citeyear{reipurth99}). Positions are derived from the {\it Spitzer}/IRAC 3.6 $\mu$m image, except SOF 2N (derived at 19 $\mu$m) and SOF 3 (derived at 8.0 $\mu$m).}
\tablehead{\colhead{SOF} & \colhead{HOPS} & \colhead{RA (2000)} & \colhead{DEC (2000)} & \colhead{IRS} & \colhead{2MASS J} & \colhead{MIR} & \colhead{FIR} & \colhead{CSO} & \colhead{VLA}}
\startdata
1  &  66     & 05:35:26.84 & -05:09:24.7 & 2       & 05352683-0509244  & 20      & \nodata & \nodata & \nodata \\
2N & 370     & 05:35:27.57 & -05:09:33.8 & 4N      & 05352762-0509337  & 21      & 3       & 22      & 11 \\
3  & 108     & 05:35:27.08 & -05:10:00.3 & \nodata & \nodata           & \nodata & 4      & 23      & 12 \\
4  & 369     & 05:35:26.98 & -05:10:17.4 & 3       & 05352696-0510173  & 27      & 5       & 24      & \nodata \\
5  & 368     & 05:35:24.72 & -05:10:29.9 & 1       & 05352477-0510296  & 28      & \nodata & \nodata & 13 \\
6  & \nodata & 05:35:26.89 & -05:11:07.6 & \nodata & 05352686-0511076  & \nodata & \nodata & \nodata & \nodata \\
7  & 60      & 05:35:23.39 & -05:12:02.6 & \nodata & 05352341-0512023  & 29      & 6b      & 25      & \nodata \\
8  & 59      & 05:35:20.14 & -05:13:15.7 & \nodata & 05352014-0513156  & 31+32   & 6d      & 30      & \nodata \\
\enddata
\end{deluxetable}

\newpage
\begin{deluxetable}{lcccccccccccc}
\setlength{\tabcolsep}{0.02in}
\rotate
\tablewidth{0pt}
\tabletypesize{\tiny}
\tablecaption{\label{tab:fluxes}Flux densities and uncertainties in Janskies for all sources detected by SOFIA/FORCAST\tablenotemark{a}.}
\tablehead{\colhead{Source} & \colhead{$F_{3.6}$} & \colhead{$F_{4.5}$} & \colhead{$F_{5.8}$} & \colhead{$F_{8.0}$} & \colhead{$F_{19.7}$} & \colhead{$F_{24}$} & \colhead{$F_{31.4}$} & \colhead{$F_{37.1}$} & \colhead{$F_{70}$} & \colhead{$F_{160}$} & \colhead{$F_{350}$} & \colhead{$F_{870}$}}
\startdata
SOF 1  & $   0.532 \pm    0.001$ & $   0.879 \pm    0.002$ & $   2.321 \pm    0.003$ & $   3.059 \pm    0.004$ & $   4.86 \pm    0.97$ & $   5.31 \pm    0.26$ & \nodata & $   4.48 \pm    0.90$ & $  \le 59.5 \pm    3.0$ & $ \le 202 \pm   10$ & $\le 12.8 \pm 5.1$ & $\le 2.9 \pm 1.2$  \\
SOF 2N & $   0.330 \pm    0.002$ & $   0.829 \pm    0.005$ & $   0.831 \pm    0.005$ & \nodata & $  15.0 \pm    3.0$ & $  45.24 \pm   10.42$ & \nodata & $ 227 \pm   45$ & $ 718.1 \pm    8.1$ & $ 482 \pm   17$ & $  26.7 \pm   10.6$ & $   \le 5.6 \pm    2.2$  \\
SOF 3  & \nodata & \nodata & \nodata & $   0.030 \pm    0.001$ & \nodata & $   1.519 \pm    0.031$ & \nodata & $   8.4 \pm    1.7$ & $ 121.2 \pm    4.9$ & $ 468 \pm   35$ & $\le   44.5 \pm    17.6$ & $ \le 7.9 \pm 3.2$  \\
SOF 4  & $   1.304 \pm    0.002$ & $   2.145 \pm    0.004$ & $   3.958 \pm    0.004$ & $   5.844 \pm    0.006$ & $  10.8 \pm    2.2$ & $  17.51 \pm    0.66$ & \nodata & $  13.9 \pm    2.8$ & $  48.0 \pm    3.2$ & $ 206 \pm   23$ & $  20.0 \pm    8.0$ & $   \le 4.8 \pm    1.9$  \\
SOF 5  & $   0.330 \pm    0.002$ & $   0.921 \pm    0.002$ & $   1.423 \pm    0.003$ & $   2.486 \pm    0.003$ & $  19.1 \pm    3.8$ & $  25.58 \pm    0.17$ & \nodata & $  65 \pm   13$ & $ 104.8 \pm    4.0$ & $ 105 \pm    9$ & $   5.9 \pm    2.4$ & $   \le 1.41 \pm    0.56$  \\
SOF 6  & $   1.345 \pm    0.001$ & $   1.476 \pm    0.002$ & $   1.457 \pm    0.001$ & $   1.490 \pm    0.005$ & $   1.24 \pm    0.25$ & $   1.291 \pm    0.010$ & \nodata & \nodata & \nodata & \nodata & \nodata & \nodata  \\
SOF 7  & $   0.018 \pm    0.001$ & $   0.047 \pm    0.001$ & $   0.044 \pm    0.001$ & $   0.044 \pm    0.003$ & \nodata & $   1.79 \pm    0.01$ & $   6.9 \pm    1.4$ & $  12.3 \pm    2.5$ & $  64.8 \pm    1.5$ & $  89.9 \pm   10.1$ & $  10.0 \pm    4.0$ & $ \le  2.14 \pm    0.86$  \\
SOF 8  & $   0.968 \pm    0.001$ & $   2.890 \pm    0.004$ & $   4.034 \pm    0.001$ & $   5.040 \pm    0.004$ & $  10.5 \pm    2.1$ & $  11.2 \pm    0.1$ & $  16.3 \pm    3.3$ & $  23.5 \pm    4.7$ & $  55.4 \pm    1.7$ & $  69.7 \pm   10.8$ & $  4.3 \pm    1.7$ & $\le 0.88 \pm 0.35$  \\
\enddata
\tablenotetext{a}{See also \citet{peterson08} for {\it Spitzer} photometry of sources detected at 24 $\mu$m.}
\end{deluxetable}

\newpage
\begin{deluxetable}{lcccccrccc@{\extracolsep{0.4em}}cc}
\setlength{\tabcolsep}{0.02in}
\tablewidth{0pt}
\tabletypesize{\small}
\tablecaption{\label{tab:modelpars}Constrained model parameters for
FORCAST-detected OMC-2 sources.\tablenotemark{a}}
\tablehead{
&&&\multicolumn{7}{c}{Envelope} &\multicolumn{2}{c}{Disk}
\\
\cline{4-10} \cline{11-12}
\\
\colhead{Source}
& \colhead{$L $}
& \colhead{$i$}
& \colhead{$\eta$}
& \colhead{$R_c$}
& \colhead{$\theta$}
& \colhead{$R_{\rm env}$}
& \colhead{$\rho_{\rm 1\,AU}$\tablenotemark{b}}
& \colhead{$\dot M_{\rm inf}$\tablenotemark{b, c}}
& \colhead{$M_{\rm env}$}
& \colhead{$R_{\rm disk}$}
& \colhead{$M_{\rm disk}$}
\\
& \colhead{($L_\odot$)}
& \colhead{(degrees)}
&
& \colhead{(AU)}
& \colhead{(degrees)}
& \colhead{(AU)}
& \colhead{($10^{-13}$ g cm$^{-3}$)}
& \colhead{($M_\odot$ yr$^{-1}$)}
& \colhead{($M_\odot$)}
& \colhead{(AU)}
& \colhead{($M_\odot$)}}
\startdata
SOF  1    & 60  & 40 & \nodata & \nodata & \nodata  & \nodata & \nodata  &
\nodata & \nodata & 40         & 1 \\
SOF 2N & 300  & 80  & 3  & 300    & 40 & 5000  & 1.5 &
$\sim$3$\times$$10^{-5}$ &  0.8     & 280    & 1.6  \\
SOF 3  & 50   & 50  &  2  & 380   & 8  & 5000  & 20 &
$\sim$4$\times$$10^{-4}$ & 10 & 380    & 0.6 \\
SOF 3\tablenotemark{d}   &  30  & 70  &  2  & 380   & 8  & 5000  & 5 &
$\sim$1$\times$$10^{-4}$ & 2.5 & 380  & 0.6  \\
SOF 4\tablenotemark{e}  & 20   & 70  & 2.5 & 100   & 0  & 5000  & 9.0  &
$\sim$2$\times$$10^{-4}$    &  4 &  \nodata   & \nodata \\
       & 60   & 40  & \nodata&\nodata&\nodata&\nodata&\nodata & \nodata &
\nodata &40 &1.5\\
SOF 5  & 40   & 50  & 2.5  & 100 & 5  & 10000 & 1.5   &
$\sim$3$\times$$10^{-5}$ & 2 & 100 & 0.2\\
SOF 7  & 15   & 70  & 2.5 & 100 & 30  & 5000  & 2.0   &
$\sim$4$\times$$10^{-5}$ & 1 &100 & 0.8 \\
SOF 8\tablenotemark{e}  &  7 & 50 & 2 & 100 &  0    & 5000  & 6.0 &
$\sim$1$\times$$10^{-4}$ & 3 & \nodata  &  \nodata \\
            & 130 & 30 & \nodata & \nodata & \nodata  & \nodata & \nodata
& \nodata  & \nodata & 100   & 0.1 \\
\enddata
 \tablenotetext{a}{$L$ is the total luminosity of the source, $i$ is the
inclination angle of the polar (rotational) axis of the system (that is
measured with respect to the line of sight), $\eta$ is a measure of the
degree of flattening of the envelope ($1\la \eta \la 4$, with $\eta = 4$
corresponding to a very flattened envelope), $R_c$ is the centrifugal
radius, $\theta$ is the aperture angle of the
envelope cavity, $R_{\rm env}$ is the outer radius of the envelope,
$\rho_{\rm 1\,AU}$ is a reference density (corresponding to the density
at a radius of 1 AU of a spherically symmetric free-falling envelope
with the same mass infall rate), $\dot M_{\rm inf}$ is
the mass infall rate (see note c), $M_{\rm env}$ is the envelope mass
(obtained by integration of the density distribution), $R_{\rm disk}$ is
the outer radius of the disk, and $M_{\rm disk}$ is the disk mass.}
 \tablenotetext{b}{Values are likely upper limits because of possible
contamination by nearby sources of FIR/submm flux densities.}
\tablenotetext{c}{Estimated from eq. 3 of \citet{kenyon93}, assuming a
nominal value of 1 $M_\odot$ for the stellar mass. $M_{\rm inf}$ is
proportional to $\rho_{\rm 1\,AU}$ and to the square root of the stellar
mass, which is not well-constrained. Values are accurate within a factor
of two for stellar masses in the range 0.25-4 $M_\odot$.}
\tablenotetext{d}{Fit obtained when $F_{160}$ is taken as an upper limit.} 
\tablenotetext{e}{Modeled as a double source. In one of the sources the
emission is dominated by a star+envelope and in the other source it is
dominated by a star+disk.}
 \end{deluxetable}

\end{document}